\documentclass[aps,pre,twocolumn]{revtex4-1}
\usepackage{amsmath}
\usepackage{amssymb}
\usepackage{appendix}
\usepackage{graphicx}
\usepackage{xcolor}
\usepackage{dcolumn}
\usepackage{bm}
\usepackage{array}
\usepackage{lipsum}

\begin{document}
	
	
	\title{Turing-like patterns in an asymmetric dynamic Ising model}
	\author{M\'elody Merle}
	\email{merle@lptmc.jussieu.fr}
	\affiliation{Sorbonne Universit\'e, CNRS, Laboratoire de Physique Th\'eorique de la Mati\`ere Condens\'ee, LPTMC, F-75005 Paris, France}
	\author{Laura Messio}%
	\email{messio@lptmc.jussieu.fr}
	\affiliation{Sorbonne Universit\'e, CNRS, Laboratoire de Physique Th\'eorique de la Mati\`ere Condens\'ee, LPTMC, F-75005 Paris, France}
	\author{Julien Mozziconacci}%
	\email{mozziconacci@lptmc.jussieu.fr}
	\affiliation{Sorbonne Universit\'e, CNRS, Laboratoire de Physique Th\'eorique de la Mati\`ere Condens\'ee, LPTMC, F-75005 Paris, France}

	\date{\today}

	\begin{abstract}

	To investigate novel aspects of pattern formation in spin systems, we  \textcolor{black}{use} a mapping between reactive concentrations in a reaction-diffusion system and spin orientations in a dynamic multiple-spin Ising model. \textcolor{black}{While pattern formation in Ising models always rely on infinite-range interactions, this mapping allows us to design an finite-range interactions Ising model that can produce patterns observed in reaction diffusion systems including Turing patterns with a tunable typical length scale. This model has asymmetric interactions and several spin types coexisting at a site. While we use the example of genetic regulation during embryo-genesis to build our model, it can be used to study the behavior of other complex systems of interacting agents.}	
	

	\end{abstract}
	
	\pacs{Valid PACS appear here}
	\maketitle
	
	\section{Introduction}
	The mechanisms underlying global pattern formation from local interactions are studied in many fields of physics, chemistry and biology \cite{castets1990experimental, koch1994biological,Diego2018}. \textcolor{black}{Reaction diffusion is among possibly the most widespread patterning model. It uses continuous variables such as the concentrations of chemical species that can form complex patterns: e.g. stripes, propagating fronts and oscillatory waves} \cite{Turing1952}. 
	\textcolor{black}{Another class of pattern formation models are Ising models. 
	Ising variables are two-states (up and down) discrete spins interacting on a lattice. One can distinguish two types of interactions: infinite and finite-range. 
	Infinite-range are more commonly referred to as long-range interactions, i.e. decreasing with the distance as a power law $1/r^\alpha$ with $\alpha$ larger or equal to the spatial dimension. Infinite and finite-range interactions show differences in terms of behavior and are investigated using specific computational methods \cite{long_range}. 
	In terms of patterning, finite-range interactions mostly lead to ferromagnetic (all spins up or down) or anti-ferromagnetic (alternating up and down spins) states. In the presence of frustration, one-site width stripes can also occur \cite{PhysRevB.87.144406}.
	On the other hand, infinite-range interactions can lead to the formation of more complex patterns. The combination of ferromagnetic nearest neighbor interactions with anti-ferromagnetic infinite-range interactions found for instance in ultra-thin magnetic films \cite{PhysRevE.86.011103, RevModPhys.72.225} is known to produce patterns such as mazes or bubbles. Up to now, infinite-range }interactions were considered as a necessary condition for the formation of patterns with a typical width of several lattice constants.
	\textcolor{black}{Our goal in the present paper is to derive a variant of the Ising model which produces such patterns without infinite-range interactions. }
	
	\textcolor{black}{Reaction-diffusion and Ising models are both well-adapted to describe emergent collective behaviors from individual evolution rules. Yet, they have been confronted in few restricted cases. An example is the rigorous proof of the existence of travelling fronts in a ferromagnetic Ising model under Glauber dynamics using a mapping of the magnetization evolution equation with the reaction diffusion Allan-Cahn equation \cite{DeMasi1994, DeMasi1995}.}
	
	In this paper, we build upon the similarities between reaction-diffusion and Ising models to construct a dynamic finite-range Ising model giving rise to complex patterns.
    We introduce in Sec.~\ref{sec:SRDA} a Stochastic Reaction Diffusion Automaton (SRDA) on a lattice, inspired by embryogenesis (Sec.~\ref{sec:aside}). Then we construct in Sec.~\ref{sec:ADIM} our Asymmetric Dynamic Ising Model (ADIM) and establish a mapping between the ADIM and the SRDA in Sec.~\ref{sec:comp}. 
    From this mapping we show in Sec.~\ref{sec:turing} that the ADIM can reproduce the variety of patterns observed in reaction-diffusion systems even if it only relies on \textcolor{black}{finite}-range interactions.

	\section{An aside on embryogenesis}
	\label{sec:aside}
	
	\textcolor{black}{One of the most emblematic example of pattern formation is embryogenesis, which motivated Turing to develop his well known reaction-diffusion model in 1952 \cite{Turing1952}.} 
    \textcolor{black}{The major question in this field is to understand how initially identical cells can express different genes to form various tissues and organs at the proper time and location during the development of the embryo. }
    

    \textcolor{black}{Turing proposed that interactions between diffusing gene products (called here after species) could lead to the spatial organization of gene expression. In the case of the so-called Turing patterns, two species (an activator $a$ and an inhibitor $b$) can self-organize in stripes, dots, or maze-like patterns. This is driven by three main ingredients:}  an auto-catalysis of $a$, an asymmetry in reciprocal interactions (i.e. $a$ enhances the formation of $b$ while $b$ penalizes the formation of $a$), and a quantitative difference in their diffusion coefficients.
	Besides these \textit{self-organized} patterns \cite{cooke1976clock, corson2017self, cotterell2015local} suggested by Turing, embryo-genesis provides examples of \textit{externally-driven} patterns that are pre-formed at large scales, for instance by chemical gradients in the case of the Drosophila's fertilized egg \cite{Wolpert1969,gregor2007stability}. These gradients activate specific genes in different locations of the embryo and the products of these genes subsequently act on other genes in a combinatorial manner.
	In the current view, the spatial regulation of gene expression in developing embryos is explained by combinations of these two different and complementary classes of scenarios \cite{Green2015}. 
	
	Various models implementing these two classes have been proposed, either based on differential equations or on cellular automata \cite{sharpe2017computer}. 
	Recently, an Ising model \cite{Hillenbrand2016a} has also been developed to model gene patterning during embryo-genesis. 
	Each spin corresponds to a gene that can be in one of the two states: active or inactive. Each site represent a nucleus and can contain several spins/genes. The spatial proximity between nuclei defines the interaction lattice (non zero couplings), while coupling values defines the interaction network between spin/genes. These couplings are a simplification of the molecular mechanisms at work: gene transcription in RNA, RNA translation in proteins and protein diffusion to neighboring nuclei where they can modify the transcription rate of other genes. While being different from more detailed reaction-diffusion models, this Ising model features the most important ingredient: short range interactions between spins/genes  \cite{Hillenbrand2016a}. Based on this similarity, we also placed ourselves in this context to build an extended \textcolor{black}{variant of the Ising model}, implementing asymmetric interactions between spin types, that could lead to Turing patterns.
	\textcolor{black}{To achieve this goal, we start by constructing a reference reaction diffusion model in the next section.}
	
	\begin{figure}
	\includegraphics[width=\linewidth]{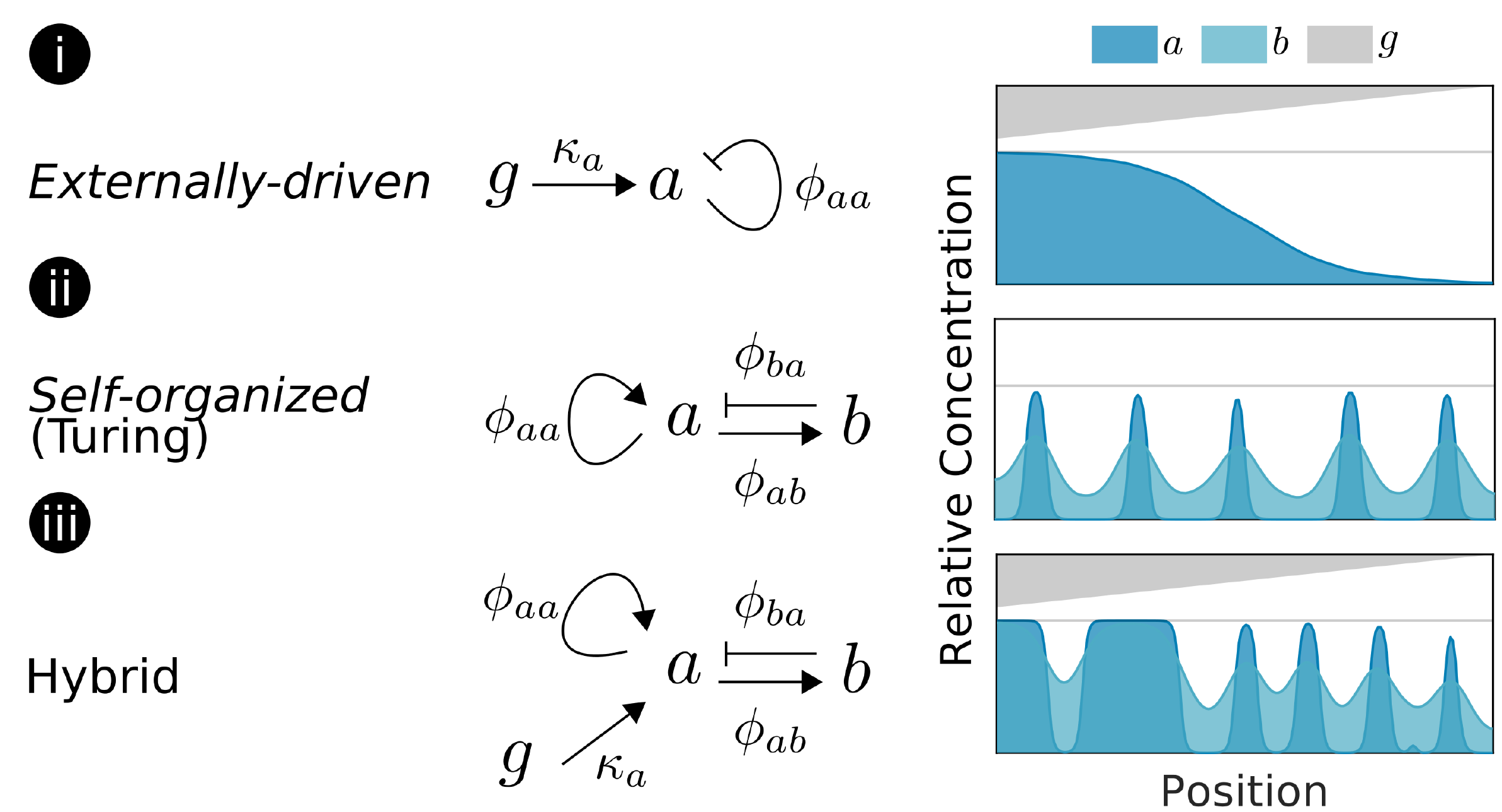}
	\caption{The SRDA recapitulates different patterning mechanisms, illustrated here on a 1D lattice.
	(i) \textit{Externally-driven} pattern by a linear external gradient $g$. (ii) \textit{Self-organized} pattern, here Turing stripes. (iii) Hybrid pattern which is a combination of both mechanisms. Corresponding interaction networks are shown with $\rightarrow$ and $\dashv$ symbols representing respectively activation and inhibition.}
	\label{fig:rdn}
	\end{figure}

	\section{Stochastic Reaction Diffusion Automaton (SRDA)}
	\label{sec:SRDA}
	
	Reaction diffusion models describe the evolution in time and space of the concentrations $\mathbf{c}=(c_a,c_b,\dots)$  of $n_s$ chemical species $a$, $b$,\dots undergoing two processes: diffusion, associated to diffusion coefficients $D_a, D_b,\dots$, and local reactions $R_a(\mathbf{c}, g), R_b(\mathbf{c},g),\dots$ resulting in the creation of  units of $a,b,\dots$ per time unit, where $g$ is a possible external gradient forcing the system. The continuous space and time equations are for each specie $a$ of the form:
	\begin{equation}
	\partial_{t}c_a = D_a \nabla^2 c_a + R_a(\mathbf{c},g)
	\label{eq:diffgen}
	\end{equation}

	As a discrete \textcolor{black}{space and time} model of reaction-diffusion, we use here a probabilistic automaton on a 1D lattice. The vector $\mathbf{c}_{i}(t) = (c_{i a}(t), c_{ib}(t), \dots)$ contains the concentrations of the $n_s$ considered species on each lattice site $i$ at discrete time $t$. The external gradient $g$ is a prescribed time-independent additional specie of concentration $g_i$ on site $i$.
	Each time step is divided in three sub-steps.
	The first one ($t\to t'$) is the production event: each concentration is incremented by 1 with probability:
	\begin{equation}
	P^{\rm prod}_{ia} = \frac{1} {1+ e^ {- u_{a}( \mathbf{C}_{i}, g_i) } }
	\label{rda:production}
	\end{equation}
	It is a common usage to model gene regulation by sigmoidal functions. 
	This exact form has been used for \textit{Drosophila} development modeling\cite{Ilsley2013}.
	We choose $u_a(\mathbf{c}_i, g_i)$ of the form:
	\begin{equation}
	u_a(\mathbf{c}_i, g_i) = \sum_{b} (\phi_{ba} c_{ib}) +\kappa_a g_i -\theta_0
	\end{equation}
	where the $\phi_{ba}$ are the reaction constants between species (the sum is over all species, including $a$ itself), 
	$\kappa_a$ represents the effects of the gradient $g$ on specie $a$ and $\theta_0$ is an activation threshold common to all species.
	
	The second event ($t'\to t''$) is diffusion, approximated by a Gaussian kernel $G_{\sigma_a}$ of standard deviation $\sigma_a$ and mean value 0:
	\begin{equation}
	c_{ia} (t'') = \sum_{|j-i| < 2\sigma_a} G_{\sigma_a}(|j-i|) c_{ja}(t')
	\label{rda:diffusion}
	\end{equation}

	The third event ($t''\to t+1$) is degradation, assumed to occur at the same rate $\varepsilon$ for all species:
	\begin{equation}
	c_{i a}(t+1) = \varepsilon c_{i a}(t'')
	\label{rda:degradation}
	\end{equation}where $ 0 < \varepsilon < 1 $.
	
	To represent the steady states, concentrations are normalized by $c_{\textrm{max}} = \frac{\varepsilon}{1-\varepsilon} $, so $c^*= \frac{c}{c_{\textrm{max}} }$ is comprised between 0 and 1.
	
	As described in Sc.~\ref{sec:aside}, two classes of patterns can emerge from our model : gradient-induced patterning (Fig~\ref{fig:rdn}-i) and Turing instabilities (Fig~\ref{fig:rdn}-ii). The two types of mechanisms can be combined (Fig~\ref{fig:rdn}-iii) leading to the appearance of hybrid patterns \cite{Green2015}.

	\begin{figure*}
		\centering
		\includegraphics[width=\linewidth]{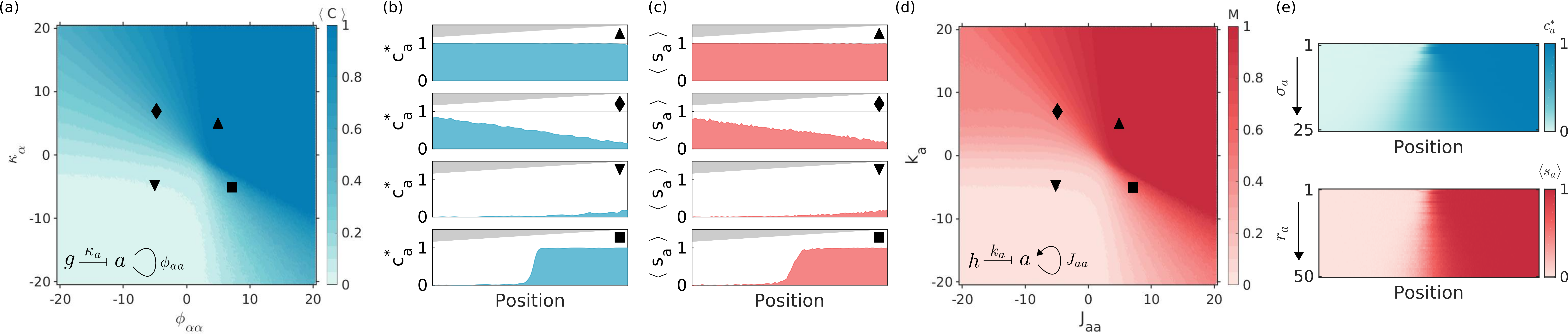}		
		\caption{Comparison of models for $n_{s}=1$, \emph{i.e.} interaction networks as represented as insets in (a,d) with 2 parameters: self-regulation $\phi_{aa}/J_{a a}$ and external gradient effect $\kappa_{a}/k_a$. Other parameters are fixed $\theta_0=h_0=1$, $\sigma_a = r_a = 1$, $\varepsilon =0.5$ and $T=1$. (a) Mean concentration and (d) magnetization in the parameters space ($\phi_{aa},\kappa_a)$ for the SRDA and ($J_{aa}, k_a)$ for the ADIM. (b) and (c) Spatial patterns obtained in both models for 4 parameters sets representing the possible types of patterns ($\blacktriangle$:fully activated, $\blacklozenge$:gradient-like, $\blacktriangledown$:fully inhibited, $\blacksquare$:sharp boundary). (d) Spatial parameters (diffusion coefficient $\sigma_a$ and interaction range $r_a$) effect on the sharpness of the $\blacksquare$-pattern. }
		\label{fig:ising}
	\end{figure*}

	\section{Asymmetric Dynamic Ising Model (ADIM)}
	\label{sec:ADIM}
	
	The classical Ising model has been created in 1920 \cite{Brush1967} as a toy model describing ferromagnetism. The spins are bivalued (usually $S_i= \pm 1$, but here, we equivalently choose $S_i=0,1$). They are placed on a lattice and interact with their nearest neighbors with an interaction constant $J>0$ that tends to align them. In the presence of a space-dependent magnetic field $h_i$, the energy of the system is:
	\begin{equation}
	\label{eq:ising}
	E = - J \sum_i\sum_{j \in \partial _{i }} { S_i \cdot S_j} - \sum_i h_{i}{S_i},
	\end{equation}
	where $ \partial _{i}$ contains the nearest neighbors of site $i$. 

	The first step to map our reaction-diffusion automaton is to design a multiple-spin Ising model, similarly to \cite{Hillenbrand2016a}. 
	To each site $i$ of the lattice, we associate $n_{s}$ spin types corresponding to the different \textcolor{black}{species} $a$, $b$, \dots. $S_i$ becomes a vector $\mathbf{S}_{\rm{i}} = (s_{i a}, s_{i b}, \dots)$ of components $0$ or $1$. 
	The concentration of the SRDA now corresponds to the average value of the spin, named magnetization.  
	Each spin type interacts with all other spin types including itself. We thus rewrite $J$ as a matrix $\bar{J}$ whose coefficient $ J_{a b}$ represents the interactions between $a$-spins and $b$-spins. To mimic the gradient effect on the different genes, $h_i$ is supposed to be coupled differently to each spin-type and we introduce a coupling vector $\mathbf{K} = ( k_{a}, k_{b}, \dots ) $, where $k_{a}$ represents the effect of $h_i$ on the $a$-spins. 
	Eq~\eqref{eq:ising} becomes:
	\begin{equation}
	E = - \sum_i\sum_{j \in \partial _{i }} \mathbf{ S_{\rm{j}}}^t \bar{J} \mathbf{S_{\rm{i}}} - \sum_i { h_i \mathbf{K} \mathbf{S_{\rm{i}}} } 
	\label{eq:isingmulti}
	\end{equation}
	
	To get patterns typical \textcolor{black}{of} reaction-diffusion, we also introduce spin type dependent interaction range $r_a$. An $a$-spin now acts on its neighboring sites up to a distance $r_a$, including itself. 
	We denote by $\partial_{ia}$ this set of spins (e.g. $\partial_{ia} = [i-r_a:i+r_a]$ on a 1D chain). 
	The interaction parameters are re-scaled using the volume $V_{\partial_{ia}}$, which is the number of sites in $\partial_{ia}$ (e.g. $V_{ \partial_{ia}} = 2r_a+1 $ in 1D). 
	
	We notice that only the average values $\frac{ J_{a b} + J_{b a}}{2}$ are important in the determination of the equilibrium state of Eq~\eqref{eq:isingmulti}, obtained at finite temperature $T \neq 0 $ using e.g. the Metropolis algorithm \cite{metropolis49}. 
	However, the matrix $\bar{J}$ needs to be effectively asymmetric to reproduce the asymmetry of the reaction constants $\phi_{ab}$ leading to the formation of interesting patterns. We thus take a step further from \cite{Hillenbrand2016a} by implementing parallel dynamics in our model, similarly to kinetic asymmetric Ising models \cite{Mezard2011, Roudi2011}, making it a non-equilibrium model.
	
	At each time step $t$, we calculate an effective field $h^{\rm{eff}}_{i a}(t)$ influencing the $s_{ia}$ spin: 
	\begin{equation}
	h^{\rm{eff}}_{i a} (t) = \sum_{b} \left( \frac{J_{b a} }{V_{\partial_{ib}}} \sum_{j\in \partial_{ib}}s_{jb} (t)\right) + k_a h_i - h_0.
	\label{eq:efffield}
	\end{equation}
	where we introduce $h_0$ an homogeneous external gradient.
	
	All spins are then updated to give the configuration at time $t+1$ according to the following probability distribution: 
	\begin{equation}
	P(s_{i a}(t+1)) = \frac{ e^{ -\beta h^{\rm{eff}}_{i a}(t) s_{i a}(t+1)}}{ 2\cosh(\beta h^{\rm{eff}}_{i a}(t))},
	\label{eq:probaising}
	\end{equation}
	where $\beta=1/T$.
	The temperature $T$ accounts for the noisiness of the system.

	\section{Model comparison}
    \label{sec:comp}
    
	The SRDA and ADIM are both governed by Markovian dynamics and share important features: asymmetry of the coupling matrices and locality of the interactions (see  App.~\ref{sec:num_methods} for the detailed numerical implementation). In the ADIM however, local reactions and transport are merged in nearest neighbors interactions. Still, many parameters appear to have similar roles: $\bar{J}$ and $\Phi$, $k$ and $\kappa$, $h_0$ and $\theta_0$, and $r$ and $\sigma$, as presented Table~\ref{table:summary}. Two parameters are nevertheless model specific: the degradation rate $\varepsilon$ and the temperature $T$. 
	
	\newcolumntype{C}[1]{>{\centering\arraybackslash}m{#1}}
	\def\arraystretch{1.5}
	\begin{table}[h!]
		\caption{Correspondence between parameters}
		\begin{ruledtabular}
			\begin{tabular}{C{3cm}|C{1cm}|C{1cm}|C{3cm}}
				\multicolumn{2}{c|}{\textbf{SRDA}} & \multicolumn{2}{c}{\textbf{ADIM}}  \\
				\hline
				Diffusion constant & $\sigma$ & $r$ & Interaction range \\
				\hline
				Reaction constant & $\bar{\phi}$ & $\bar{J} $ & Interaction strength \\
				\hline 
				Reaction constant with gradient & $\mathbf{\kappa}$ & $k$ & Coupling with linear field\\
				\hline 
				Activation threshold & $\theta_0$ & $h_0$ & Homogeneous external field\\
				\hline 
				Degradation & $\varepsilon=0.5$ & & \\
				\hline
				Noise &  & $T$=1 & Temperature\\
			\end{tabular}
		\end{ruledtabular}
		\label{table:summary}
	\end{table}

	We wish to investigate the similarity in patterns that can be obtained by confronting the two models.
	We study the case $n_s=1$, which corresponds to a classical Ising model under a space-dependent external field. We fix $T=1$ in the ADIM since this temperature gives the most direct equivalence with Eq~\eqref{rda:production} (see App.~\ref{sec:varepsilon}.). For the sake of simplicity, we fix $r_a=\sigma_a=1$ so that only between nearest neighbors interact  and we choose the space-dependent external field to be linear.
	In this conditions, we show by a calculation using the homogeneous solutions in mean-field approximation (see App.~\ref{sec:varepsilon}) that the degradation rate that gives the best mapping between interaction parameters in both models is $\varepsilon_{\textrm{opt}}=0.5$.
	
	Fig~\ref{fig:ising}-a-d presents the mean concentration $\langle c^* \rangle$ and the magnetization $M$ of the patterns obtained by the SRDA and ADIM in the planes ($\phi_{aa}$, $\kappa_a$) and ($J_{aa}$, $k_a$) with $h_0=\theta_0=1$. They represent the fraction of time during which $a$ is being produced or spins $s_a$ are up, averaged over each positions. For all probed regions of the parameter space, very similar patterns are observed for both the ADIM and the SRDA. Two broad zones of $\langle c^* \rangle=M=1$ and $\langle c^* \rangle=M=0$ correspond to full activation (Fig~\ref{fig:ising}-$\blacktriangle$) and full inhibition (Fig~\ref{fig:ising}-$\blacktriangledown$). Transition between theses two states can be done either through the formation and displacement of a sharp boundary (Fig~\ref{fig:ising}-$\blacksquare$), corresponding to strong inhibition by the external gradient and auto-activation, or by the appearance and intensification of a smooth gradient-like pattern (Fig~\ref{fig:ising}-$\blacklozenge$), corresponding to a gradient activation balanced by a null or very low auto-inhibition. 
	
	Yet, the sharpness of the $\blacksquare$-pattern is different in both models. Intuitively this sharpness depends on the correlation length of the system. The difference in patterns thus reflects a quantitative difference between the roles of the diffusion coefficient and the  interaction range. Fig~\ref{fig:ising}-e represents the evolution of the $\blacksquare$-pattern in SRDA and ADIM as a function of respectively on $\sigma_a$ and $r_a$. The maps confirm that $\sigma_a$ and $r_a$ are the spatial parameters that control the boundary sharpness. In both models, increasing these parameters causes an increase in the correlation lengths and thus a decrease of the border sharpness. It is therefore possible to choose a value for $r_a$ for which the slope exactly matches the slope obtained for any given value of $\sigma_a$ (see App.~\ref{sec:eqsigmar} for more discussion on this matter).

	\section{Turing patterns in the ADIM} 
	\label{sec:turing}
	
		\begin{figure}
		\includegraphics[width=\linewidth]{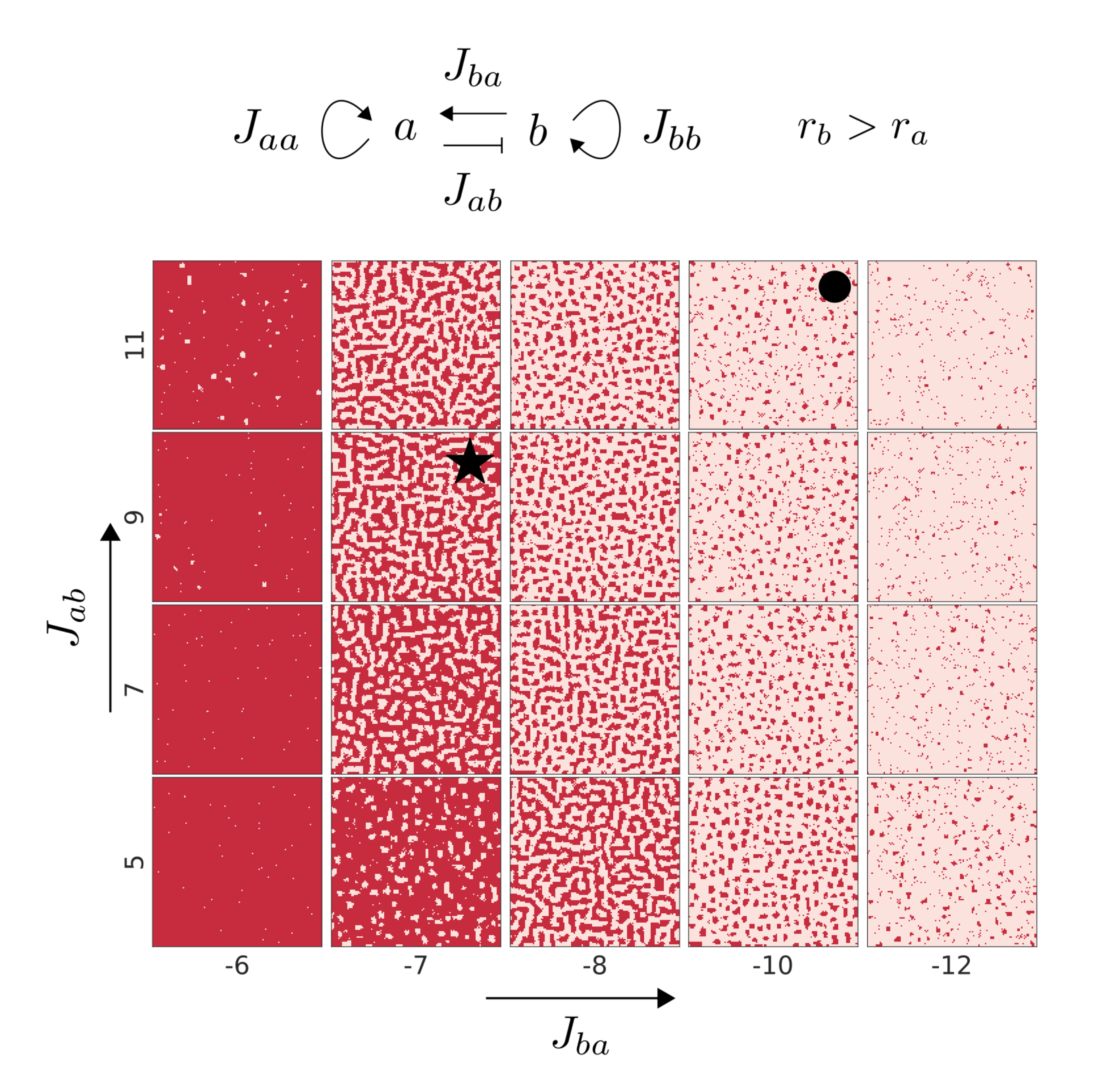}
		\caption{\textcolor{black}{ 2D ADIM simulations on 128x128 periodic square lattice with $h_{0}=1$, $r_{a}=1$, $r_{b}=5$, $J_{aa} = 13$, $J_{bb}=0, $ $J_{ab}$ and $J_{ba}$ varying, and $T=1$. Dark red corresponds to $s_a=1$ and pink to $s_a=0$. A variety of Turing-like patterns is observed including "maze" $\bigstar$ and "bubbles" $\bullet$.}}
		\label{fig:turing}
	\end{figure}
	
	In light of these similarities between SRDA and ADIM in the case $n_s=1$, we now look at the case $n_s=2$ with no external gradient and try to reproduce Turing patterns. To investigate a richer variety of Turing patterns, we also switch to 2D with periodic boundary conditions. A main feature enabling Turing instabilities in reaction diffusion models is the difference of diffusion coefficients for both species. Likewise, the ratio ${r_b}/{r_a}$, where $a$ is the activator spin-type and $b$ is the inhibitor spin-type, needs to be large enough to form patterns of different spin orientations. 2D patterns as a function of the interactions $J_{ab}>0$ and $J_{ba}<0$ are presented in Fig~\ref{fig:turing}-a for $r_a=1$ and $r_b=5$, \textcolor{black}{on a $128\times128$ square lattice.} A subtle trade-off between the values of the interactions gives rise to different types of patterns, including "maze" $\bigstar$ and "bubbles" $\bullet$. \textcolor{black}{These two patterns are used for further analysis in the next part.}

	\begin{figure*}
		\includegraphics[width=\linewidth]{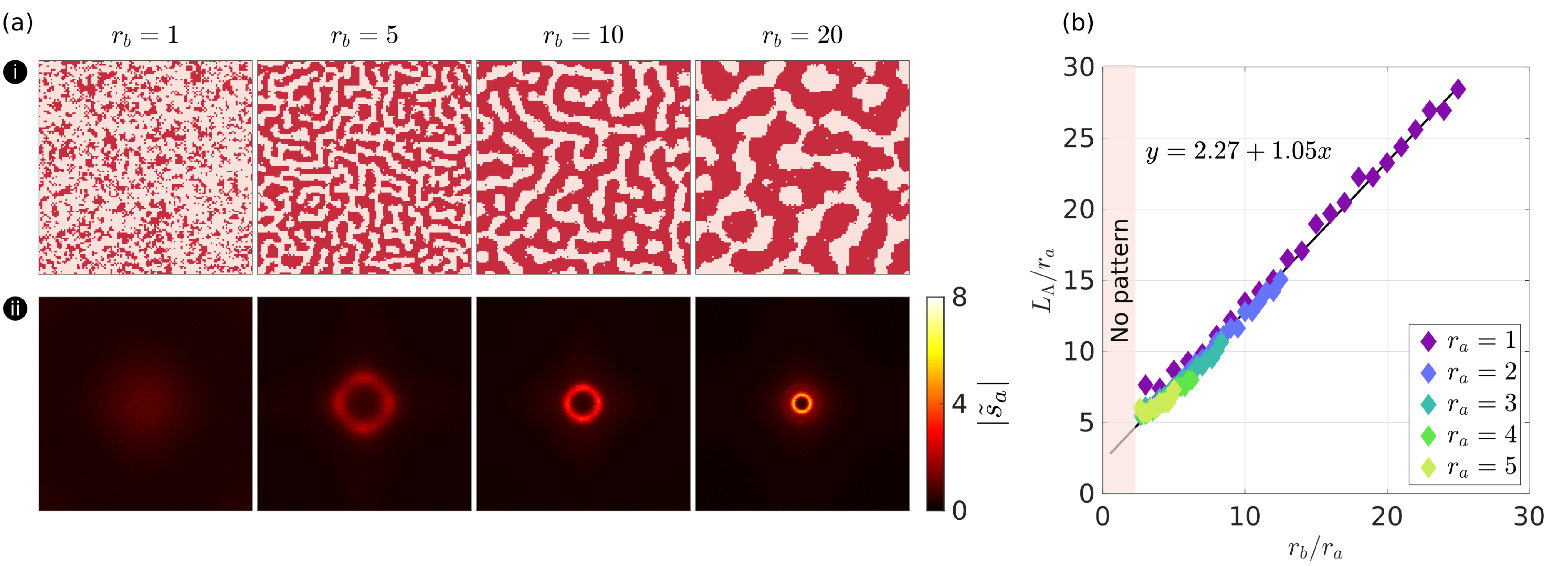}
		\caption{\textcolor{black}{(a): (i) $128 \times 128$ zoomed snapshots of the "maze" pattern obtained in simulations on $512\times512$ periodic square lattices with $r_a=1$ and $r_b=1, 5, 10, 20 $ (Dark red corresponds to $s_a=1$ and pink to $s_a=0$). (ii) The corresponding 2D Fourier Transforms averaged over $n=20$ independent simulations. (b) The value of the typical length $L_{\Lambda}/r_a$ of the patterns as a function of the ratio $r_b / r_a$ for different values of $r_a$.}}
		\label{fig:turingrange}
	\end{figure*}

    To quantitatively study the 2D patterns obtained with our ADIM, \textcolor{black}{we compute the discrete 2D Fourier Transform (FT) of $s_a$ on our periodic square lattice of linear size $n$:}
    
    \textcolor{black}{
    \begin{equation}
        \tilde s_a(q_x, q_y)= \sum_{x=0}^{n-1} \sum_{y=0}^{n-1} e^{-\frac{2i\pi}{n}(xq_x+yq_y)} s_a(x,y)
    \end{equation}
    }
    
    \textcolor{black}{ These patterns have one characteristic length resulting in a circular distribution in the reciprocal space (Fig~\ref{fig:turingrange}a). The FT gives two quantities: the pseudo order parameter $\Lambda$ and the characteristic length $L_{\Lambda}$. We average $\tilde s_a$ over circles of radius $q$ to get the mean intensity $\lambda(q)$ and define the order parameter $\Lambda = \lambda(q_{\textrm{peak}})$ as the peak value of these curves. The corresponding characteristic length in the direct space is $L_{\Lambda}= \frac{1}{q_{\textrm{peak}}}$. For illustration, Fig~\ref{fig:tf} shows the $\lambda(q)$ curves at different temperatures for both "maze" and "bubbles" patterns.}
    
    \textcolor{black}{We first investigate the role of $r_b$ in the emergence of patterns for a fixed $r_a=1$. Fig~\ref{fig:turingrange}a shows the "maze" pattern and its FT for different values of $r_b$. While there is no pattern for small values of $r_b$, maze patterns appear for higher values. Computing $L_\Lambda/r_a$ for $r_b$ ranging from 1 to 25 we conclude that the maze pattern appears for $r_b \geq 3$ (Fig~\ref{fig:turingrange}b). The characteristic length of the pattern increases linearly with $r_b$ with a coefficient close to 1.
    For higher values of $r_a$, $L_\Lambda/r_a$ increases linearly with the ratio $r_b/r_a$ with the same coefficient. We conclude that if one takes $r_a$ as a unit length, the behavior of adimensional quantities $L_\Lambda/r_a$ and $r_b/r_a$ is independent of the lattice spacing. When the lattice spacing becomes small compared to the characteristic length $r_a$ we can refine the value of the ratio $r_b/r_a$ for which patterns appear. In this case we find that this value tends towards 2.}  
	
	\begin{figure*}
	    \centering
	    \includegraphics[width=\linewidth]{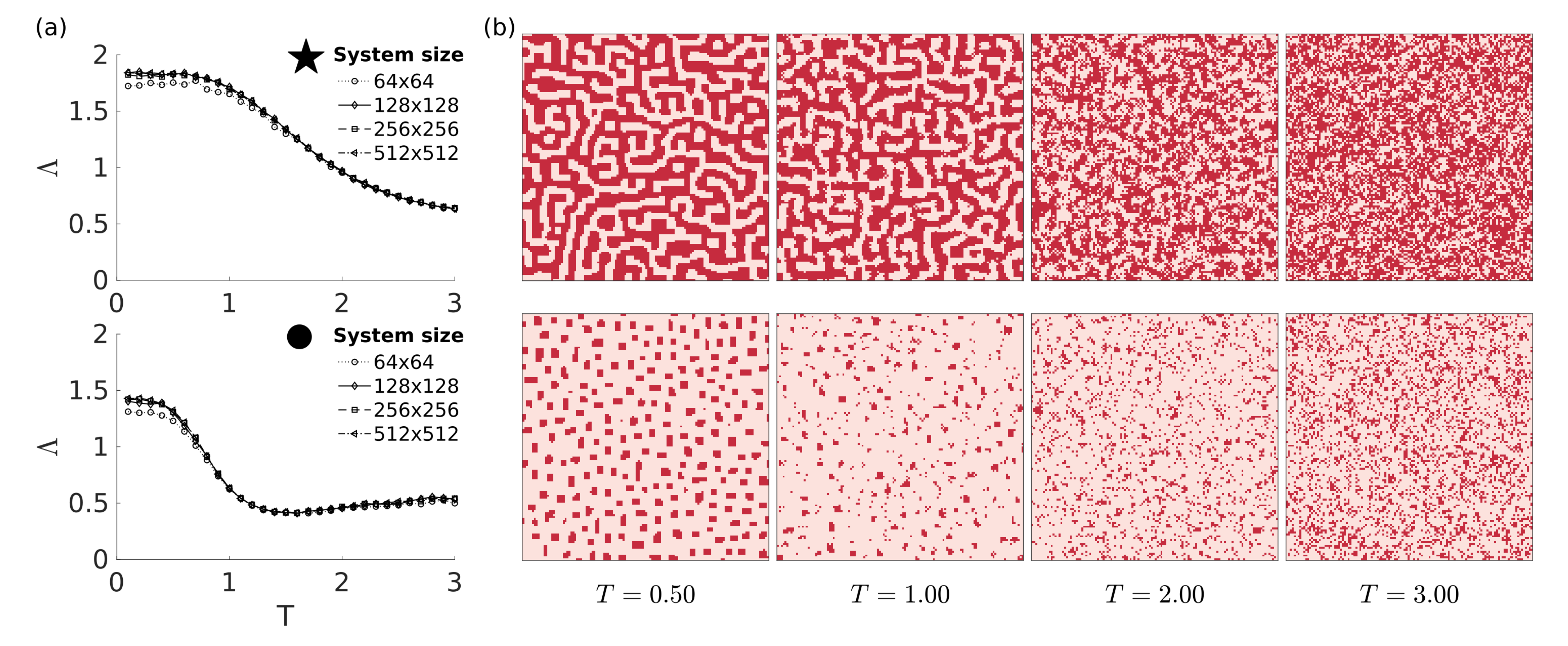}
	    \caption{\textcolor{black}{(a) Pseudo order parameter $\Lambda$ as a function of temperature for "maze" ($\bigstar$) and "bubbles ($\bullet$) patterns issued from Fig~\ref{fig:turing}. (b) Snapshots at different temperatures of the corresponding patterns. }}
	    \label{fig:phasetransition}
	\end{figure*}
	
	We next investigate the effect of temperature on the stability of these patterns \textcolor{black}{using the pseudo order parameter $\Lambda$ defined above}. In infinite-range interacting Ising models at equilibrium, phase transitions between smectic, nematic or liquid phases can occur when $T$ varies \cite{PhysRevB.51.1023}. Here, the various patterns do not break any symmetry and no such phase transition can occur. 
	Fig~\ref{fig:phasetransition} shows $\Lambda$ for different system sizes and different temperatures. A dynamic crossover occurs between a patterned and a disordered phase for both patterns, albeit at different temperatures. The crossovers do not depend on the lattice size, thus excluding potential finite size effects. 
	
    In the Ising model, the temperature reflects thermal noise, whereas in a reaction diffusion system, noise is related to the number of molecules: a lower number leads to larger relative fluctuations in the local concentrations. Molecular dynamics studies point out the importance of fluctuations for the emergence of Turing patterns \cite{Lemarchand2011}. Our results on the stability of patterns with temperature suggest on the other hand that if the number of molecules becomes too small (corresponding to high temperatures, hence high fluctuations) the patterns could disappear.

	\section{Conclusion}
	\label{sec:conclusion}
	
	To summarize, using a embryo-genesis-inspired mapping between an Ising model and a reaction-diffusion \textcolor{black}{automata}, we have constructed the ADIM, a \textcolor{black}{finite}-range out-of-equilibrium Ising model that can give rise to Turing patterns with a typical length scale of several lattice spacings. Such Turing patterns had previously only been observed in \textcolor{black}{infinite}-range Ising models. 
	\textcolor{black}{It is worth noticing that our variant of the Ising model can be seen as a cellular automata featuring rules similar to the ones which can be derived from reaction-diffusion equations (see for instance \cite{Weimar1994}). Our approach shows that a discrete toy model with different ranges of interaction and an asymmetry in the interaction coupling between species is sufficient to produce Turing-like patterns.}

	\renewcommand{\thefigure}{S\arabic{figure}}
	\setcounter{figure}{0}
	\renewcommand{\theequation}{S\arabic{equation}}
	\setcounter{equation}{0}
	
	\twocolumngrid

        \appendix
	
	\section{Numerical methods}
	\label{sec:num_methods}
	
	\textit{ADIM}--- 
	The system is composed of $N_S$ sites. On each site is associated $n_s$ spin types. 
	Parallel dynamics is implemented as follows.
	
	We initialize the system at $t=0$ by randomly attributing a value $0$ or $1$ at each of the $n_s \times N_S$ spins.  At each time step $t$ we calculate the effective field $h^{\rm{eff}}_{i a} (t)$ for each $a$-type spin on each site $i$:
	\begin{equation}
	h^{\rm{eff}}_{i a} (t)= \sum_{b} \left( \frac{J_{b a}}{V_{\partial_{ib}}} \sum_{j\in \partial_{ib}}s_{jb}(t) \right) + k_a h_i - h_0,
	\label{eq:efffieldsup}
	\end{equation}
	where $J_{ba} $ is the action of $b$-spins on $a$-spins, $h_i$ the value of the space dependent external magnetic field at site $i$, $k_a$ the coupling between $\mathbf{h}$ and $a$-type spins and $h_0$ a global external field.
	$\partial_{ib}$ designates the nearest neighbors of spin $s_{ib}$, defined by the interaction range $r_b$ for $b$-spin type such as $\partial_{ib} = \{j, |j-i| < r_b  \}$ and $V_{\partial{ib}} =Card(\partial_{ib})$. 
	
	We calculate the corresponding Boltzmann probabilities:
	\begin{equation}
	\label{eq:probaisingsup}
	P(s_{i a}(t+1) =1) = \frac{ e^{ \beta h^{\rm{eff}}_{i a}(t)}}{ 2\cosh(\beta h^{\rm{eff}}_{i a}(t))}
	\end{equation}
	where $\beta=T^{-1}$ is the inverse temperature. 
	
	In 1D, the simulation runs at a fixed temperature $T$. In 2D, the stationary state is calculated using a simulated annealing (SA) to avoid freezing in configurations with long relaxation times.Cooling strategy is exponential:
	\begin{equation}
	T_{SA}(t) = T_0\times a^t
	\end{equation}
	with $0<a<1$.
	
	The spins at time $t+1$ are simultaneously chosen according to these probabilities.
	The operation is repeated during $n_t$ time steps, and in case of SA, $a$ is chosen so that $T_{SA}(n_t) = T$, \textit{ie} $a=\big( \frac{T}{T_0} \big) ^{n_t^{-1}}$. Then the simulation is pursued  and recorded for $n_\textrm{rec}$ time steps. Finally $\langle s_{ia} \rangle$ is obtained by averaging over the $n_\textrm{rec}$ recorded configurations. 
	Typically, $n_t=250$ and $n_\textrm{rec}=50$.

    \begin{figure*}
    \centering
	     \includegraphics[width=\linewidth]{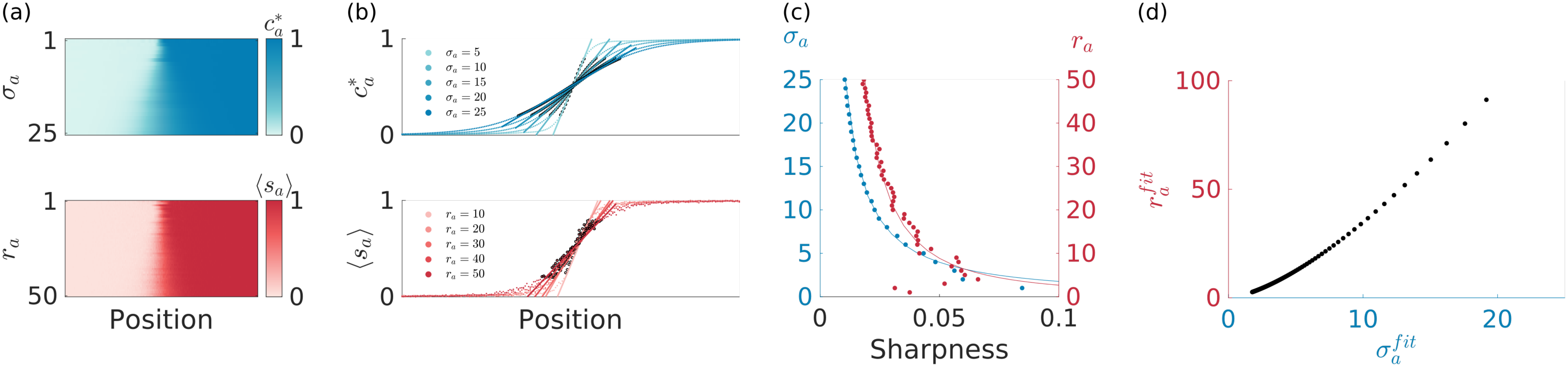}
	     \caption{Correspondence between $r_a$ using the sharp-boundary pattern introduced in main text (see Fig~\ref{fig:ising}). (a) Increasing the spatial parameters $\sigma_a$ and $r_a$ decreases the boundary sharpness. (b) Quantitative evaluation of the sharpness by fitting the boundary interface by an aff
	     ine function. (c) $\sigma_a$ and $r_a$ as a function of the measured sharpness are fitted by power laws $y=\frac{a}{x^b}$. For any given sharpness, $\sigma_a^\textrm{fit}$ and $r_a^\textrm{fit}$ can be calculated using these fits. (d)  $r_a^\textrm{fit}$ as a function of $\sigma_a^\textrm{fit}$.}
	     \label{fig:apprange}
	 \end{figure*}

	\textit{SRDA}---
	The system is composed of $N_S$ sites at which is associated $n_s$ specie concentrations $c_a$. 
	Each concentration is randomly chosen at $t=0$ between $0$ and $1$. Each time step is subdivided in 3 events as presented in the main text. 
	
	This is repeated for $n_t$ time steps to reach the stationary state. Then each concentration is normalized by $c_\textrm{max}$ to obtain $ 0 < c^*_{ia} < 1$. Typically, $n_t= 200 $. 

	In both models, the boundary conditions are set to reflecting in the case 1D to have gradient-dependent patterns. In the 2D ADIM, we choose periodic boundary conditions. 
	
	\section{Determination of $\varepsilon_{\textrm{opt}}$ using homogeneous stationary solutions}
	\label{sec:varepsilon}

	We use here the homogeneous steady state solutions in mean-field approximation for both models to establish a relationship between parameters in the case of one specie $a$ ($ n_s=1$), and more specifically to find the optimal value $\varepsilon_{\textrm{opt}}$ for which both model lead to the same patterns for equal network parameters. 
	
	\textit{ADIM}---
	The mean-field approximation for the ADIM is (see Eq.\eqref{eq:probaising}):
	\begin{equation}
	\langle s_{ia} \rangle	 = \frac{1}{1 + e^ {\beta \langle h^\textrm{eff}_{ia} \rangle } }
	\label{eq:isingmf}
	\end{equation}
	For only one spin type $a$ ($n_s=1$):
	\begin{equation}
	\langle s_{ia} \rangle	 = \frac{1}{1 + e^ {\beta \big[ \frac{J_{aa}}{V_{\partial_{ia}}} \sum_{j\in \partial_{ia}} \langle s_{ja} \rangle + k_a h_i - h_0 \big] } }
	\end{equation}
	Since we consider a homogeneous state and we are in 1D,  $\sum_{j\in \partial_{ia}} \langle s_{ja} \rangle = (2r_a+1) \langle s_{ia} \rangle$ and $V_{\partial_{ia}} = 2r_a +1$, leading to:
	\begin{equation}
	\langle s_{ia} \rangle	 = \frac{1}{1 + e^ {\beta \big[ J_{aa}\langle s_{ia} \rangle + k_a h_i - h_0 \big] } }
	\end{equation}

	\textit{SRDA}---
	In the mean-field approximation for the SRDA model, we average over several realizations of time evolution for each time step. Thus, we replace the probabilistic increment of the production step (Eq~\eqref{rda:production}) by a deterministic one: 
	\begin{equation}
	c_{i a}(t') = c_{i a}(t) + \frac{1}{1+ e^{-u_a(\mathbf{c}_{i}, g_i)}}.
	\end{equation}
	By combining this new equation to the two other events (diffusion (Eq~\eqref{rda:diffusion}) and degradation (Eq~\eqref{rda:degradation}) ), we get:
	\begin{equation}
	c_{i a}(t+1) =  \varepsilon \bigg[ \sum_{|j-i| < 2\sigma_a} G_{\sigma_a}(|j-i|) [  c_{j a} (t) + \frac{1}{1+ e^{-u_a(\mathbf{c}_{j})  } } ] \bigg]
	\label{eq:mfrda}
	\end{equation}

	In the classical PDE reaction-diffusion approach as defined in Eq~\eqref{eq:diffgen}, searching for an homogeneous steady state equates to look for the solution of $R_a(\mathbf{c},g)=0 $ since $\partial_t c_a =0$ (steady state) and $\nabla^2c_a=0$ (homogeneous). In our SRDA, the homogeneity translates in $\forall j$ $c_{ja}(t) = c_{ia}(t)$, and using the fact that $\sum_{r<2\sigma} G_\sigma(r) = 1 $, we obtain: 

    \begin{equation}
    \sum_{|j-i| < 2\sigma_a} G_{\sigma_a}(|j-i|) [  c_{j a} (t) + \frac{1}{1+ e^{-u_a(\mathbf{c}_{j})  } }] = c_{ia}(t')
    \end{equation}
	which leads to a simplification of Eq~\eqref{eq:mfrda}  :
	\begin{equation}
	c_{i a}(t+1) = \varepsilon \bigg[ c_{i a} (t) + \frac{1}{1+ e^{-u_a(\mathbf{c}_{i},g_i)}} \bigg]
	\end{equation}
	
	Being at steady-state $c_{ia}(t+1) = c_{ia}(t)$, and using $c_{max} = \frac{\varepsilon }{1 - \varepsilon} $, we retrieve $c^*$ :	
	\begin{equation}
	c^*_{i a} = \frac{1}{1+ e^{-u_a(\mathbf{C}_{i})}} 
	\end{equation}
	For $n_s=1$:
	\begin{equation}
	c^*_{i a} = \frac{1}{ 1+ e^{\big[\phi_{aa} \frac{c^*_{ia} (1-\varepsilon) }{\varepsilon}+ \kappa_a g_i - \theta_0 \big]} } 
	\label{eq:rdnhomog}
	\end{equation}

	\textit{Optimal mapping value of $\varepsilon$}---
	Eq~\eqref{eq:rdnhomog} and Eq~\eqref{eq:isingmf} are combined to search for the solution of $\langle s_{ia} \rangle = c^*_{ia}$, leading to:
	\begin{align}
	\beta 
	\big[ J_{aa}\langle s_{ja} \rangle + k_a h_i - h_0 \big] 
	= \frac{\phi_{aa} \varepsilon}{1-\varepsilon} c^*_{ia} + \kappa_a g_i - \theta_0.
	\end{align}
	
	For $\beta=T^{-1}=1$, identical network parameters ($J_{aa}=\phi_{aa}$, $k_a=\kappa_a$ and $h_0 = \theta_0$) and under the same external gradient ($h_i=g_i$), we get:
	\begin{equation}
	\langle s_{ia} \rangle = \frac{\varepsilon}{1-\varepsilon} c^*_{ia}.
	\end{equation}
	Finally, we obtain the optimal degradation parameter $\varepsilon$ for which both models gives equivalent homogeneous steady-state for equivalent parameters, as presented in the results of the main content of this letter Fig~\ref{fig:ising} :
	\begin{equation}
	\varepsilon = 0.5
	\end{equation}

		    \begin{figure*}
		\includegraphics[width=\linewidth]{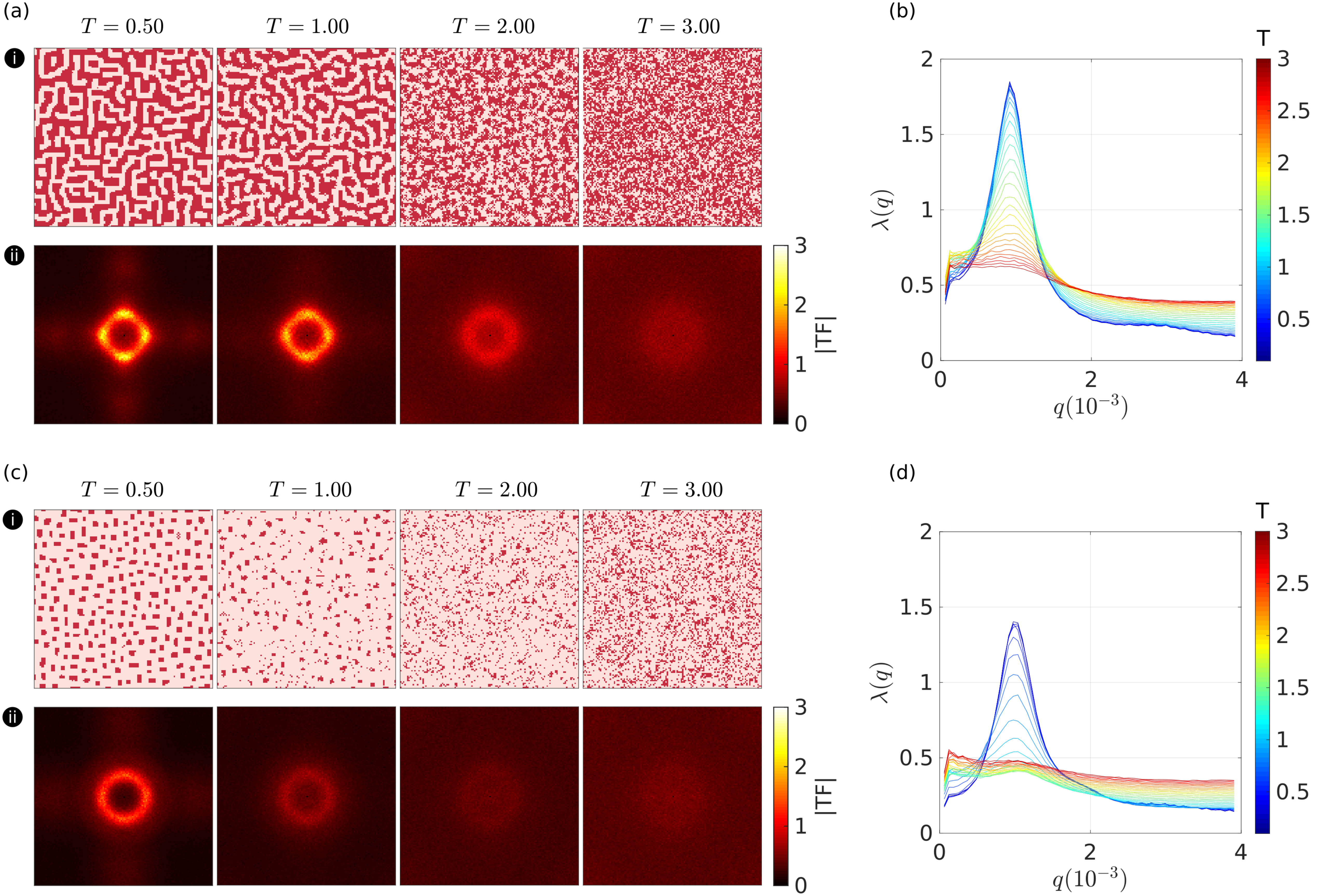}
		\caption{(a), (c): (i) Snapshots of patterns in $128\times128$ periodic square boxes for $T = 0.5, 1,2,3$ for respectively "maze" and "bubbles" patterns introduced in main text (Dark red corresponds to $s_a=1$ and pink to $s_a=0$) and (ii) 2D Fourier transforms of theses patterns averaged on $n=100$ independent simulations. (b), (d): Mean value $\lambda(q)$ of the FT over circles of radius $q$ over different temperatures between 0 and 3. The peak value of this distribution gives us the order parameter $\Lambda$. The corresponding length in direct space $L_\Lambda$ is the characteristic length of the pattern.}
		\label{fig:tf}
	\end{figure*}

	 \section{Equivalence between $\sigma_a$ and $r_a$}
	 \label{sec:eqsigmar}

	 To evaluate the correspondence between $\sigma_a$ in the SRDA and $r_a$ in the ADIM we use the sharp boundary pattern presented in the main text for $n_s=1$ under a linear external gradient. This pattern appear when the spin/gene activates itself but is repressed by the gradient ($J_{aa} =7$ and $k_a=-5$).
	 As seen in the main text and shown again in Fig~\ref{fig:apprange}-a, as $\sigma_a$ or $r_a$ increases, the sharpness of the boundary decreases. We first measure this sharpness by fitting the boundary interface ($0.2 < c_a^*, \langle s_a \rangle < 0.8$ ) by a affine function of the position (Fig~\ref{fig:apprange}-b). The slope defines the sharpness of the boundary. Both the diffusion constant $\sigma_a$ and the interaction range $r_a$ as a function of sharpness are fitted by power laws $y=\frac{a}{x^b}$ (Fig~\ref{fig:apprange}-c). From these fits we calculated $\sigma_a^{\textrm{fit}}$ and $r_a^{\textrm{fit}}$ for any given sharpness and plotting $r_a^\textrm{fit}$ as a function of $\sigma_a^\textrm{fit}$ gives us a equivalence between $r_a$ and $\sigma_a$ (Fig~\ref{fig:apprange}-d).


	\bibliographystyle{apsrev4-1}
	\bibliography{main.bib}

\end{document}